\documentclass[mathleft
]{an}
\usepackage{graphicx}
\usepackage{times}
\usepackage{url}
 \usepackage{amssymb}
\begin{document}

\Pagespan{1}{}
\Yearpublication{2015}%
\Yearsubmission{2015}%
\Month{}%
\Volume{}%
\Issue{}%

\title{TANAMI -- \\
       Multiwavelength and Multimessenger Observations of Active Galaxies}

\author{M. Kadler\inst{1}\fnmsep\thanks{\email{matthias.kadler@astro.uni-wuerzburg.de}\newline}
\and 
R. Ojha\inst{2,3,4}
on behalf of the TANAMI Collaboration
}
\titlerunning{TANAMI}
\authorrunning{M. Kadler \& R. Ojha}
\institute{
Institut f\"ur Theoretische Physik und Astrophysik, Emil-Fischer Str. 31, 
D-97074 W\"urzburg, Germany
\and
NASA, Goddard Space Flight Center, Greenbelt, MD 20771, USA
\and
Catholic University of America, Washington, DC 20064, USA
\and
University of Maryland, Baltimore County, Baltimore, MD 21250, USA
}

\received{}
\accepted{}
\publonline{}


\abstract{%
Extragalactic jets launched from the immediate vicinity of supermassive black holes in radio-loud active galactic nuclei (AGN) are key objects in modern astronomy and astroparticle physics. AGN jets carry a fraction of the total gravitational energy released during the accretion of matter onto supermassive black holes and are prime suspects as possible sources of ultrahigh-energy cosmic rays and the recently detected extraterrestrial neutrinos at PeV energies. TANAMI (Tracking Active galactic Nuclei with Austral Milliarcsecond Interferometry) is a multiwavelength program monitoring AGN jets of the southern sky. It combines high-resolution imaging and spectral monitoring at radio wavelengths with higher-frequency observations at IR, optical/UV, X-ray and $\gamma$-ray energies. We review recent results of the TANAMI program, highlighting AGN candidate neutrino-emitters in the error circles of the IceCube PeV neutrino events.
  }

\maketitle

\section{Introduction}
Active Galactic Nuclei (AGN) are, by far, the largest category of \textsl{identified} sources in the \textsl{Fermi}/LAT Third Source Catalog (3FGL: \cite{3fgl}, 3LAC: \cite{3lac}), with blazars being the dominant source type. This confirmed the strong link between the physics of blazars and $\gamma$-ray emission, first suggested based on data from the 
\textit{Compton Gamma Ray Observatory} with its EGRET detector (\cite{Tho93,Don95}).
Blazars are copious emitters of highly variable emission in every waveband, forming the most \textsl{active} subset of AGN. High resolution observations of blazars show relativistic collimated outflows of plasma from a center of activity which is associated with a supermassive central object (\cite{Beg84}). These jets emit brightly across the electromagnetic spectrum, often exhibit apparent superluminal motion (\cite{Coh07}), and are involved in the regulation of star formation and galaxy evolution via AGN feedback (\cite{McN07}). 

Observations and modeling of blazar phenomena now span about half a century and impressive progress has been made in understanding many aspects of these mysterious objects. However, the fundamental questions regarding their composition, formation, collimation and dissipation are still open. There are several reasons for this of which the biggest is their variability. This creates three challenging demands on observations that seek to address blazar physics: they must be (i) frequent, (ii) broadband, and (iii) simultaneous. The launch of the {\it Fermi} satellite (in June 2008), commissioning of large TeV telescopes, the availability of X-ray satellites, access to the 100\,GHz through 1\,THz band with ALMA, combined with a host of optical/NIR and radio facilities, have made it possible for the first time to make observations that are well constrained throughout the electromagnetic spectrum and can be made quasi-simultaneously (see, e.g., \cite{Ojh13}). 

High resolution radio observations have a unique role in unraveling the physics of blazars that goes beyond their important role in constraining the low energy end of the spectral energy distribution (SED) in at least two vital ways. First, VLBI (Very Long Baseline Interferometry) monitoring of AGN provides the only direct measure of relativistic motion in AGN jets, probing intrinsic jet parameters such as speed, Doppler factor, opening and inclination angles. Second, the unmatched resolution of VLBI makes identifying the location and extent of emission regions possible through observed changes in both VLBI morphology and flux, correlated with changes in the high energy emission. 

In combination with observations at other wavelengths, VLBI observations are addressing some of the most fundamental questions posed by the detection of $\gamma$-ray emission from blazars:
Where are the $\gamma$-rays produced in blazar jets and what is the extent of these emission regions? 
What are jets made of? Which physical processes regulate the high-energy emission of blazars? Are AGN also responsible for the production of ultrahigh-energy cosmic rays? Here, we describe a multiwavelength program monitoring AGN in the southern sky, which is also taking advantage of new opportunities in multimessenger astrophysics.

\section{The TANAMI Multiwavelength Program}
It has long been apparent that broadband, quasi-simultaneous data on a large sample of AGN in different activity states are needed to address the key questions mentioned above. In anticipation of the launch of the {\it Fermi Gamma-ray Space Telescope}, TANAMI began observing an initial sample of 43 AGN in 2007. Addition of subsequent {\it Fermi}-LAT detections continues to expand the sample which currently contains more than 90 monitored sources. Starting as a radio program with its focus on high-resolution VLBI observations and flux-density monitoring, the program has developed into a multiwavelength program providing excellent NIR/optical/UV, X-ray and $\gamma$-ray coverage for a large sample of $\gamma$-ray loud AGN. 

The unique role of parsec scale radio monitoring of $\gamma$-ray AGN is well appreciated (\cite{Jor11,Lis11}). However, most monitoring programs use northern hemisphere arrays that cannot observe the far southern sky. Since the beginning of the \textsl{Fermi} mission, TANAMI is the only large VLBI monitoring program targeting AGN south of declination $-30^{\circ}$ (\cite{Ojh10}). Uniquely among large VLBI programs, TANAMI observes at two frequencies, 8.4 and 22.3\,GHz, measuring spectral indices of jet features and their time evolution.
TANAMI uses a southern-hemisphere VLBI network (\cite{Pre89}) consisting of the five telescopes that make up the Australian Long Baseline Array (LBA), an antenna at Hartebeesthoek, South Africa, and the 34\,m and 70\,m telescopes of the NASA Deep Space Network in Tidbinbilla. Through a program approved by the International VLBI Service, these are typically joined by two German antennas, GARS at O'Higgins, Antarctica and TIGO which was at Concepcion, Chile, but which has now been moved to La Plata, Argentina. Lately, the fidelity of the TANAMI array got an additional boost with access to 
two AuScope antennas in Yarragadee, Western Australia and Katherine, Northern Territory (\cite{Lov13}) 
and both a telescope in Warkworth, New Zealand and a single ASKAP antenna also in Western Australia
(\cite{Tzi10}).

A radio flux-density monitoring program  with the ATCA (Australia Telescope Compact Array) at frequencies of 4.8, 8.6, 17/19, 38 and 40\,GHz (\cite{Ste12}) and a single-dish monitoring program using the University of Tasmania telescope at Ceduna at 6.7\,GHz
are supplementing the VLBI observations.
A single-baseline interferometer, the Ceduna Hobart Interferometer (CHI) formed with the radio telescopes in Hobart, Tasmania, and Ceduna, South Australia, can respond to a \textsl{Fermi} trigger within half an hour (\cite{Bla12a}). 

To cover the
NIR/optical part of the spectrum, 22 TANAMI sources were monitored
through 2011/2012 with the Rapid Eye Mount (REM) telescope in J, H, K,
and R bands (\cite{Nes13}). Additional REM time was granted in
2014 and monitoring of another 6 sources is in progress. Many sources
that did not have measured redshifts, or even an optical
identification, were observed using Gemini South and are now being
targeted using the TNG (Telescopio Nazionale Galileo) telescope on La
Palma. Many TANAMI sources are being observed with the Nordic Optical
Telescope, the SARA consortium telescope on CTIO, and ATOM (for some
sources). We also take advantage of public data from the SMARTS
program. Coverage in the UV band is provided by the UVOT instrument on
board \textit{Swift} (see below).

Over many years, the TANAMI sample has had an unusually uniform X-ray coverage. Due to the lack of a sensitive enough X-ray all-sky monitoring instrument, this has been obtained via many pointed 
observations via the guest-observer programs of various X-ray telescopes: 60\,ksec on \textit{RXTE}, 260\,ksec on \textit{XMM-Newton}, 125\,ksec with \textit{Swift}, 800\,ksec with \textit{INTEGRAL}, and 70\,ksec with \textit{Suzaku}. 
In addition, a total of 300\,ksec of optical/UV and X-ray observing time comes from a dedicated \textit{Swift} fill-in program. 

The \textit{Fermi} Large Area Telescope (LAT, \cite{Atw09}) has been monitoring the sky at 30\,MeV to 300\,GeV since its launch in 2008. 
A large number of AGN were detected and monitored with \textit{Fermi}/LAT (709 AGN during the first year of operation, \cite{1lac}; 886 AGN after two years, \cite{2lac}; 1563 AGN after four years, \cite{3lac}).

In recent years, TANAMI has established a close liaison with the ANTARES collaboration. ANTARES is an undersea neutrino telescope in the Mediterranean Sea off the coast of Toulon, France (\cite{antares}), completed in 2008 (coincident with the launch of \textit{Fermi}). 885 photomultiplier tubes mounted on twelve vertical lines are anchored to the seabed at a depth of $2475$\,m and register Cherenkov light induced by relativistic muons produced in charged-current interactions of high-energy neutrinos with matter around the detector. 
ANTARES is optimised for up-going muons produced by neutrinos from southern-hemisphere sources which have traversed the Earth, and achieves its highest sensitivity at declinations south of about $-30^\circ$, coinciding with the TANAMI sky. 

\section{Selected Results}
\subsection{The (Sub-)Parsec-Scale Structure and Dynamics of AGN Jets} 
First epoch VLBI images of the initial TANAMI sample of 43 objects at 8.4\,GHz included many sources observed for the first time at milliarcsecond resolution (\cite{Ojh10}). These images revealed morphological differences between $\gamma$-ray bright and faint sources with the brighter jets having larger projected opening angles possibly indicative of smaller angles to the line of sight.
This was also
shown independently, based on data from
the MOJAVE program (\cite{Pus09}). VLBI images of the extended TANAMI
sample have been published as part of several TANAMI studies (\cite{Mue11,Bla12a,Mue13,Mue14,Mue14a,Mue14b,Kra14,Kra15}) or as supplementary data in multiwavelength studies of the \textit{Fermi}/LAT collaboration (\cite{Abd09a,Abd10c,Abd10d,Bus14}). Data from the TANAMI ATCA program were used in \textit{Fermi}/LAT multiwavelength publications on the TANAMI sources PKS~B\,2123$-$463 and PKS~B\,2142$-$758 (\cite{dAm12,Dut13}). VLBI images of both of these sources will be presented as part of a comprehensive catalog of 8.4\,GHz VLBI images of all TANAMI sources observed with VLBI beyond the initial sample (M\"uller et al., in prep.).

\begin{figure}
\includegraphics[width=\columnwidth]{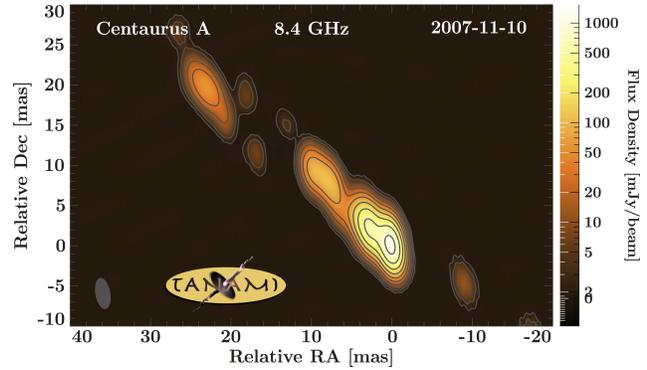}
\caption{\label{fig:cena} First frame of a movie, available via {\tt http://www.aip.de/AN/movies}, showing the structural evolution
of the inner 60\,mas of the Cen\,A jet-counterjet system. 
The movie is based on eight images obtained by TANAMI
between 2007 and 2011, which were convolved with a common beam as indicated by
the grey ellipse in the bottom left corner and interpolated between the individual 
observations. The color scale was fixed to the
minimum and maximum flux density of the images. The contour lines start at
three times the maximum noise level of the images and increase
logarithmically by factors of two.
}
\end{figure}

The jet-counterjet system of the closest radio-loud active galaxy Centaurus A (Cen\,A), at a distance of only 3.8\,Mpc, has been studied within the TANAMI VLBI program on unprecedented small linear scales (\cite{Mue14b}). The high linear-resolution observations provide a unique view of the inner parsec of an AGN jet. A $3.5$\,yr kinematic study showed the evolution of the central-parsec jet structure of Cen\,A, 
revealing 
complex dynamics  indicative of a spine-sheath jet
structure and downstream acceleration of jet plasma (see Fig.~\ref{fig:cena}). Both moving and stationary jet features are found. A persistent local minimum in surface brightness suggests the presence of an obstacle interrupting the jet flow, which can be explained by the interaction of the jet with a star at a distance of $\sim0.4$\,pc from the central black hole. 

\subsection{High-Energy Emission of Blazars and Other Classes of AGN}
The TANAMI sample was originally defined as a hybrid radio and $\gamma$-ray selected sample that includes most radio-loud extragalactic jets south of $-30^\circ$ that had either been detected at $\gamma$-ray energies during the EGRET era or were considered candidates for $\gamma$-ray emission. Indeed, the majority of TANAMI sources have now been detected in $\gamma$-rays after more than six years of \textit{Fermi}/LAT observations, either during flares or over long-time integration. 
The largest group of non-detected TANAMI sources are radio galaxies, but 
also some radio-bright quasars remain notoriously $\gamma$-quiet. Due to source variability, the best statistical results can be obtained from quasi-simultaneous data sets and relatively short time averages. We have performed a radio-gamma correlated analysis of TANAMI sources (\cite{Boe12,Boe15}) based on the 11-month 1FGL period (\cite{1fgl}). A large fraction (72\,\%) of all TANAMI sources is associated with $\gamma$-ray sources for this time range. Association rates differ for different optical classes with all BL Lacs, 76\,\% of quasars and 17\,\% of the TANAMI galaxies being detected by \textit{Fermi}/LAT. We also calculated upper limits on the $\gamma$-ray flux for $\gamma$-faint sources. We found a correlation between the $\gamma$-ray and radio luminosities with $L_\gamma \propto L_\textrm{VLBI}^{0.90\pm0.04}$ and showed that this correlation is not induced by the common distance-squared term in both quantities. 
Some of the upper limits are still statistically compatible with this correlation but we also find a group of apparently intrinsically $\gamma$-ray faint quasars. We calculated brightness temperatures of the radio cores during the 1FGL period and showed that they correlate with $\gamma$-ray luminosity. Brightness temperatures of several sources above the inverse Compton limit imply strong Doppler boosting, which is in general agreement with other studies (e.g., \cite{Kov09,Lin11}).

The vast majority of AGN detected by \textit{Fermi}/LAT are blazars. In addition to those, new classes of non-blazar $\gamma$-ray bright AGN have been discovered by the LAT. The elusive class of $\gamma$-ray detected \textsl{misaligned} radio-loud AGN has grown to a size of 29 sources (\cite{Abd10e,3lac}).
Apart from three radio galaxies (Cen\,A, Cen\,B, and Pic\,A), the source PKS~B\,0625$-$354 is in the TANAMI sample. While showing an extended FR\,1 structure (\cite{Mor99}), the optical spectrum of PKS~B\,0625$-$354 resembles that of a BL\,Lac object (\cite{Wil04}). TANAMI observations revealed a single-sided, parsec-scale jet morphology (\cite{Mue13}) with a high-brightness-temperature core, suggesting that the $\gamma$-ray emission of this object is produced in a similar way as in blazars. 
A new class of $\gamma$-ray emitting AGN detected by \textit{Fermi}/LAT are radio-loud Narrow-Line Seyfert\,1 Galaxies (RLNL-Sy\,1s, \cite{Abd09b,Abd09c}). VLBI observations of PMN~J\,0948+0022 (\cite{Abd09d,Gir11}) have demonstrated that this source has a high-brightness-temperature radio core, suggesting that the $\gamma$-ray emission of \textit{Fermi}/LAT-detected RLNL-Sy\,1s is coupled to the presence of relativistically 
beamed jets as in
blazars, but harboured in
spiral rather than elliptical galaxies and originating from the environment of supermassive black holes with relatively low-masses (\cite{Fos13}). 
Among the five $\gamma$-ray detected RLNL-Sy\,1s, the radio-loudest one, PKS\,2004$-$447, is located in the southern hemisphere and exhibits peculiar properties in this elusive sample. TANAMI VLBI images (Schulz et al., in prep.) show an extended, but diffuse one-sided jet with a dominating unresolved VLBI core. Along with its radio spectral classification as a CSS galaxy (\cite{Gal06}) and its compact structure on arcsec scales, this might suggest that PKS\,2004$-$447 is a young jet source pointed at a relatively small angle to the line of sight. This is also in agreement with new TANAMI X-ray observations with \textit{XMM-Newton}, showing a  blazar-like X-ray spectrum with only marginal evidence for the presence of the previously reported soft excess (\cite{Kre13}, Kreikenbohm et al., in prep.). 

An unusual $\gamma$-ray source detected by {\it Fermi}-LAT is PMN~J\,1603$-$4904 which was originally classified as a BL Lac object. Imaged at VLBI resolution for the first time by TANAMI, it shows a hard to classify morphology; it is either an atypical blazar or a $\gamma$-ray loud young radio galaxy (\cite{Mue14a}). If confirmed as the latter, this would be the first of this class of objects to be detected in $\gamma$-rays. Subsequently, near simultaneous observations with \textit{XMM-Newton} and \textit{Suzaku} detected a redshifted X-ray line (\cite{Mue15a}) making it unlikely to be a blazar and thus supporting the possibility that PMN~J\,1603$-$4904 is a young radio galaxy with a misaligned jet. 

\subsection{SED Modeling}  
The SED of blazars have two maxima. Synchrotron emission from relativistic electrons in the jet produces the low energy peak in the radio-to-infrared band (\cite{Bla79}). The origin of the high energy component, which lies in the MeV to TeV bands, remains an open question. It could arise from inverse-Compton upscattering of synchrotron photons by the electrons which emitted them (Synchrotron Self Compton model; e.g., \cite{Koe81}) or from the inverse-Compton scattering of photons external to the jet by the relativistic electrons within the jet (External Compton model; e.g., \cite{Sik94}). Yet another class of models suggest this second component arises from hadronic processes involving high-energy protons which produce neutral and charged pions that decay into $\gamma$-ray photons and neutrinos (e.g., \cite{Man93}). Since blazar emission is variable at all wavelengths and the relationships between these variations is unclear, multiepoch, quasi-simultaneous observations across the spectrum are essential to distinguish between these models. 

TANAMI is playing a major role, with studies of a number of blazars in different states of $\gamma$-ray activity (\cite{Bla12b,Bus14,Dut13,Bus14,Car15}). Flaring and quiescent states were modeled with a leptonic model (based on \cite{Fin08} and \cite{Der09}) that includes the synchrotron, synchrotron self-Compton (SSC), and the external Compton (EC) processes. 
Fits to the SED were first attempted with a single seed photon source of EC and a broken power-law electron distribution but in some cases the data required more complicated models introducing more free parameters, e.g., changing the electron distribution from a broken power-law to a double broken power-law. 

The resulting models show a wide variety in the fitted parameters even for different flares in the same blazar (\cite{Cha13}), so at first glance each flare appears to be different. Fortunately, even with a modest (but increasing) number of modeled flares a tentative pattern is emerging with two types of flares apparent. 
Type\,I flares require changes in the electron distribution to explain the observed changes in their SED between active and quiescent $\gamma$-ray states, e.g., PKS\,0537$-$441 (\cite{dAm13}).
On the other hand, Type\,II flares require changes in additional parameters such as the magnetic field or the size of the emitting region (or both) , e.g., flare B in PKS\,2142-758 (\cite{Dut13}).
Both Type\,I and Type\,II flares can be sub-divided into two types. In the case of Type\,Ia flares, the $\gamma$-ray flare is accompanied by an optical flare. For Type\,Ib flares the optical flare is either weak or absent. Type\,IIa $\gamma$-ray flares are accompanied by a weak or absent X-ray flare
whereas Type\,IIb flares are accompanied by a strong X-ray flare.
More observations over flaring and quiescent states are needed to establish or improve this tentative classification scheme which has the potential to bring order to our study of high energy blazar emission.

\subsection{Neutrino Emission of Radio-Loud AGN}
In SED analysis, an unambiguous separation of the emission components due to the relativistic electrons, positrons, protons, or ions is often not possible due to the superposition of a large number of emission zones within the jets. Even the superb broadband SEDs that can be collected since the beginning of the
\textit{Fermi}/LAT mission do not allow us to unambiguously discriminate  
between hadronic and leptonic emission processes, and to pin down the matter content of AGN jets, 
which have long been {suspected as the origin}  of ultrahigh-energy cosmic rays (\cite{Hil84}). 
Secondary neutrino emission would then naturally explain a high-energy neutrino flux
exceeding the atmospheric background above energies of 100\,TeV (\cite{Man92,Man95}). A tentative association of high-energy neutrinos with flaring blazars have been suggested (\cite{Hal05, ANTARES_Flare}) but to date
no unambiguous identification of an AGN as a neutrino source could be established. 

The major challenge for neutrino telescopes is to detect a weak signal in the presence of a strong background of atmospheric muons and neutrinos, which are not directly associated with any cosmic sources. One way to increase the significance is to search for time-clustered multiple events from the same direction and/or the search for correlations with flaring activity in the electromagnetic spectrum. ANTARES has performed a search for time-clustered neutrino events from eight TANAMI sources (\cite{Fri14}). For the source PKS\,0208$-$512, two events were found, for PKS\,1954-388 one event, which however is not a significant signal. In another joint analysis,  \textit{Fermi}/LAT light curves are used to select suitiable time windows during short flares for an ANTARES analysis (\cite{Mue14,Feh15}), which can enhance the sensitivity of the detector substantially (see also \cite{ANTARES_Flare}).

Recently, the IceCube collaboration has reported the detection of three neutrinos at PeV energies and a larger number of sub-PeV events in excess of the atmospheric background (\cite{Ice1,Ice2,Ice3}).
Due to the steeply falling background towards higher energies, the PeV events are the most likely ones in this sample of events to be of extraterrestrial origin.
Using contemporaneous TANAMI multiwavelength data, we investigated the six radio- and $\gamma$-ray brightest blazars located in the median angular resolution uncertainty
circles reported by the IceCube collaboration for the
first two PeV neutrino events, dubbed `Ernie' and `Bert' (\cite{Kra14}).
Our VLBI images revealed typical blazar morphologies for most candidate sources
with high-brightness-temperature radio cores and typically one-sided jets indicative of strong differential Doppler boosting. PKS\,B0302$-$623 shows a highly peculiar halo-like morphology
and PMN\,J1717$-$3342 is found to be affected by strong scatter broadening. Broadband SED analysis revealed
prominent blue bumps of Swift~J1656.3$-$3302 and PMN\,J1717$-$3342, suggesting that the latter's earlier classification as a BL\,Lac object should be revised.

Interpreting the high-energy emission  of the six candidate blazars in a hadronic scenario, in which accelerated protons give rise to photoproduction of pions and subsequent $\gamma$-ray and neutrino emission, the neutrino luminosity of these sources can be estimated from the observed flux of the high-energy bump in the SED. 
We could demonstrate that  the integral energy output of the six TANAMI candidate blazars is high enough to explain the observed neutrino flux. No single blazar in the field, however, yields a high-enough calorimetric output for an unambiguous association with the highest predicted neutrino fluxes being ascribed to Swift~J1656.3$-$3302 and PMN\,J1717$-$3342. Instead, we showed that the larger population of faint unresolved blazars can contribute a comparable fraction to the total neutrino flux in a field of several degrees in size
(\cite{Kra15}) so that a direct association with an individual source might currently only be possible in exceptional cases.

The ANTARES and TANAMI collaborations have followed up on these results by analysing the sub-PeV neutrino emission of the six TANAMI candidate blazars (\cite{A+T15}). Although much smaller in volume, ANTARES offers better sensitivity below $\sim 100$\,TeV at the given southern declinations. A standard point-source analysis revealed two signal-like events, one fitted to the position of Swift~J1656.3$-$3302 and the second fitted to the position of PMN\,J1717$-$3342, the same two sources which showed the highest predicted neutrino output. Still,
an atmospheric origin cannot be excluded. No ANTARES events are observed from any of the other four blazars, constraining at a 90\,\% confidence level the possible neutrino power-law spectral indices to be flatter than $−2.4$. 

An analysis of the TANAMI AGN in the field of the third PeV neutrino event detected by IceCube (dubbed `Big Bird') is currently ongoing. The field holds two remarkable objects: the high-fluence outburst blazar PKS\,1424$-$418 and the closest radio-loud AGN to Earth, Cen\,A. 
In the latter case, the possible jet-star interaction in the inner jet of Cen\,A (see above) provides an interesting scenario for a possible source of the neutrino emission. We investigated the possibility  that neutrinos are produced in the 
interaction of protons accelerated at a bow shock surrounding a red giant star and its wind in the jet
(\cite{Mue15b}) but found that an association with the Big Bird neutrino event is highly improbable due to the low predicted neutrino flux.

\section{Outlook}
Quasi-simultaneous, multiwavelength, multiepoch observations of the TANAMI program are providing answers to many conundrums about $\gamma$-ray loud AGN. With up to seven years of VLBI monitoring, most of the original 43 sources of the 2007 TANAMI pilot program have reached the stage at which robust studies of jet kinematics are now yielding a rich harvest of results. Further TANAMI epochs will allow us to determine basic kinematics for sources added more recently to the sample. Equally important, they are needed to determine component ejection epochs which can be correlated with  $\gamma$-ray flares or  neutrino emission. 

To understand the energetics of blazars it is imperative to probe them in different states of activity by establishing a sample of sources with simultaneous observations in every waveband. This can only be achieved by long-term monitoring programs such as TANAMI lasting several years. The new availability of data from ALMA and \textit{NuSTAR}, at previously inaccessible bands, increases the value of new observations. TANAMI observations will also image structural changes in high brightness sources targeted in snapshot observations by the RadioAstron radio satellite. TANAMI is uniquely well positioned to exploit multimessenger observations which have long been anticipated and are now a reality. 

\acknowledgements
The authors thank the rest of the TANAMI team for their manifold contributions to the overall program and for advise and comments on the manuscript.
The Australia Long Baseline Array is part of the Australia Telescope National Facility which is funded by the Commonwealth of Australia for operation as a National Facility managed by CSIRO.
This study made use of data collected through the
AuScope initiative. AuScope Ltd is funded under the National
Collaborative Research Infrastructure Strategy (NCRIS), an Australian
Commonwealth Government Programme.

\end{document}